\documentclass[fleqn,10pt]{wlscirep}
\usepackage[utf8]{inputenc}
\usepackage[T1]{fontenc}
\usepackage{upgreek}
\usepackage{tabularx}
\usepackage{import}
\usepackage{adjustbox}
\usepackage{mathtools}
\usepackage{float}
\usepackage{graphicx}
\usepackage{textcomp}
\usepackage{gensymb}
\usepackage{booktabs}
\usepackage{multirow}
\usepackage{siunitx}
\usepackage{pifont}
\usepackage{amsmath}
\usepackage{amssymb}
\usepackage{verbatim}
\usepackage{hyperref}
\usepackage{scalerel,stackengine,amsmath}
\newcommand\equalhat{\mathrel{\stackon[1.5pt]{=}{\stretchto{%
    \scalerel*[\widthof{=}]{\wedge}{\rule{1ex}{3ex}}}{0.5ex}}}}
\DeclareUnicodeCharacter{25CF}{$\bullet$}

\makeatletter
\def\thickhline{%
  \noalign{\ifnum0=`}\fi\hrule \@height \thickarrayrulewidth \futurelet
   \reserved@a\@xthickhline}
\def\@xthickhline{\ifx\reserved@a\thickhline
               \vskip\doublerulesep
               \vskip-\thickarrayrulewidth
             \fi
      \ifnum0=`{\fi}}
\makeatother

\newlength{\thickarrayrulewidth}
\title{\Huge Addressing materials' microstructure diversity using transfer learning}

\author[1,2,*,+]{Aurèle Goetz}
\author[1,3,+]{Ali Riza Durmaz}
\author[4,5]{Martin Müller}
\author[1,3]{Akhil Thomas}
\author[4,5]{Dominik Britz}
\author[2,6]{Pierre Kerfriden}
\author[1,3]{Chris Eberl}
\affil[1]{Fraunhofer Institute for Mechanics of Materials IWM, Freiburg, 79108, Germany}
\affil[2]{Mines ParisTech PSL University, Centre des Matériaux, Evry, 91000, France}
\affil[3]{University of Freiburg, Freiburg, 79110, Germany}
\affil[4]{Saarland University, Department of Materials Science, Saarbrücken, 66123, Germany}
\affil[5]{Material Engineering Center Saarland, Saarbrücken, 66123, Germany}
\affil[6]{Cardiff University, School of Engineering, Cardiff, UK}

\affil[1]{ali.riza.durmaz@iwm.fraunhofer.de}
\affil[*]{aurele.goetz@mines-paristech.fr}
\affil[+]{these authors contributed equally to this work}


\begin{abstract}
Materials' microstructures are signatures of their alloying composition and processing history. Therefore, microstructures exist in a wide variety.
As materials become increasingly complex to comply with engineering demands, advanced computer vision approaches such as deep learning (DL) inevitably gain relevance for quantifying microstrucutures' constituents from micrographs. While DL can outperform classical computer vision techniques by large margins for many tasks, shortcomings are poor data efficiency and domain generalizability across data sets. This is inherently in conflict with the expense associated with annotating materials data through experts and extensive materials diversity.
\\ \\
To tackle poor domain generalizability and the lack of labeled data simultaneously, we propose to apply a sub-class of transfer learning methods called unsupervised domain adaptation (UDA). This class of learning algorithms addresses the task of finding domain-invariant features when supplied with annotated source data and unannotated target data, such that performance on the latter distribution is optimized despite the absence of annotations. Exemplarily, this study is conducted on a lath-shaped bainite segmentation task in complex phase steel micrographs. Here, the domains to bridge are selected to be different metallographic specimen preparations (surface etchings) typical for experimental scatter and distinct imaging modalities. We show that a state-of-the-art UDA approach surpasses the na\"ive application of source domain trained models on the target domain (generalization baseline) to a large extent. This holds true independent of the domain shift, despite using low-quantity data, and even when the baseline models were pre-trained or employed data augmentation. Through UDA, mean intersection over union (mIoU) was improved over generalization baselines from 82.2\%, 61.0\%, 49.7\% to 84.7\%, 67.3\%, 73.3\% on three target datasets, respectively. This underlines this techniques' potential to cope with materials variance.  
\end{abstract}
\begin{document}

\flushbottom
\maketitle
%
%
\thispagestyle{empty}


\section{Introduction}

\subsection{Motivation}
The inner structure of a material, the so-called microstructure, determines most properties and shows substantial variance depending on its composition and process history. Hence, the quantification and digitization of microstructures play an important role in virtual material design as well as objective and automated quality control. Currently, the quantification of microstructural constituents is performed on material sections, which undergo mechanical polishing and chemical etching routines to expose these inner features in e.g. light optical microscopy (LOM) or scanning electron microscopy (SEM). Aside from the material itself, this process introduces substantial scatter in the micrographs' appearance. With evolving materials complexity, not only the quantitative micrograph assessment by metallography experts becomes increasingly subjective and costly, but also conventional computer vision algorithms have reached their limits. 

Therefore, the materials science field recently adopted deep learning models for this task but is hampered by the amount of available and specifically labeled data. While annotated data is scarce, input micrographs are acquired of various materials, with different settings and optics, by different experts, and with varying processing. This motivates the need for models that can generalize across such data \textit{domains} without additional labeled data in the target domains. However, deep learning approaches have been shown to exhibit poor generalization capability across miscellaneous data sets \cite{torralba2011unbiased, welinder2013lazy, beery2018recognition} and materials science data sets \cite{Thomas2020, Durmaz2021}. Furthermore, in terms of annotated data, materials parameter spaces are often sparsely (i.e., disjointly) populated, contributing to poor model generalization. Advanced generalization techniques could enable model sharing between institutes and their applicability on diverse data sets. Microstructure characterization is one of many tasks, which could benefit from such models.

This work addresses DL-based microstructure quantification and generalization thereof while focusing on a binary segmentation (pixel-wise classification) task introduced in our last paper \cite{Durmaz2021}. It encompasses the segmentation of the lath-shaped bainite phase on topography-contrast SEM images of an etched complex phase steel surface. In this particular case, the annotation process is not only expensive but also complex, to the extent that there is a frequent disagreement between different experts on the exact nature of some phases \cite{fielding2013bainite}. This impairs the development of DL models. In our previous work \cite{Durmaz2021}, an extensive effort has been made to combine the SEM micrographs with orientation-contrast from electron back-scatter diffraction (EBSD) images, facilitating a more repeatable and precise labeling process. However, this time-consuming procedure emphasizes the demand for frugal models in terms of labeled training data.  

Moreover, this work draws the focus on model transferability to a target domain in the context of low source data availability and unavailability of target domain annotations. These are typical boundary conditions in materials science. As a first step, having trained a model on a source dataset, we want to give some insight into the possibility of adopting it to a target domain (domain generalization experiments). In this case, we investigate whether pre-training with domain-extrinsic datasets and subsequent fine-tuning to the source domain can improve generalization to target domains in the low data regime. To this end, we compare different pre-training datasets with models trained from random initialization. Subsequently, we apply a state-of-the-art unsupervised domain adaptation (UDA) model introduced by Tsai et al. \cite{tsai2020learning} to our phase segmentation task across different domains. With the aid of additional \textit{unlabeled} target data, this technique attempts to learn domain-invariant features to facilitate the domain transfer. Provided few data, and especially in the materials science field, it is unresolved whether such advanced deep learning techniques can facilitate the transfer across different processing routes or even materials. As example studies, distinct metallographic surface etchings and different imaging modalities were investigated as the domains to bridge.

\subsection{Related work}

Instances where deep learning was applied to solve materials science tasks, are fairly scarce. For instance, Azimi et al. \cite{Azimi_2018} perform a microstructure segmentation task on SEM images of dual-phase steels using a fully convolutional deep neural network coupled with a super-pixel voting approach. Holm et al. investigate various tasks, amongst others microstructure segmentation, on their ultra-high carbon steel (UHCS) dataset using an end-to-end deep learning approach \cite{decost_lei_francis_holm_2019}. Recently, Thomas et al. \cite{Thomas2020} published a study implementing a U-Net architecture for damage segmentation, detecting fatigue-induced extrusions and cracks on steel micrographs.

\subsubsection{Fine-tuning pre-trained models}
When training with little data, the first prevalent practice in DL is fine-tuning a pre-trained model\cite{Yosinski2014HowTA}. This procedure is often used synonymously with \textit{transfer learning} even though latter now encompasses many further methods. Pre-training a neural network is based on the assumption that its first layers learn similar features regardless of the trained task. Indeed, these layers act as feature extractors, detecting edges, corners, colors, or blobs. Thus, carrying over the weights of a model pre-trained on a large-scale dataset and fine-tuning them on an another task reduces the demand for training data in the latter task while accelerating convergence. Different strategies exist, ranging from full model fine-tuning with a small learning rate to freezing the initial layers of the model and fine-tuning only the last ones.

Utilizing readily available model weights from pre-training with ImageNet, owing to its apparent size and richness, has become the status quo for weight initialization in deep learning. Recently, He et al. \cite{He2019RethinkingIP} questioned the undifferentiated ImageNet weight usage, showing that it only helps in faster convergence but does not enable performance improvement. However, the low-data regime, i.e., when little data is available for the final task, was shown to be exempt from this finding. Specifically, pre-training on ImageNet culminated in improved COCO object detection \cite{lin2015microsoft} only when less than 10k images were used for fine-tuning. 

However, in material science, this low-data regime is a lasting attendant circumstance of both materials diversity and annotation complexity, making pre-training a suitable strategy to infer robustness and to raise the performance of the trained models \cite{Taj2016}. Nevertheless, pre-training dataset selection remains an unsettled issue. 

In other domains, this has been the subject of recent research \cite{CHEPLYGINA201921,Romero2020}. The overarching trends seem logical --- using a large pre-training dataset as close as possible to the target task dataset proves beneficial. However, the trade-off between data quantity and domain gap is undefined. Cheplygina et al. \cite{CHEPLYGINA201921} condensed 12 studies from the medical field in a detailed review, comparing ImageNet with smaller in-field datasets. The conclusions are not consistent, and no clear trend could be identified. Romero et al. \cite{Romero2020} studied the effect of different pre-training for chest X-ray radiographs classification, varying the number of target training images from 50 to 2,000. Pre-training was conducted using ImageNet or one of two X-ray datasets (220k chest images and 40k images of various body parts). They show that pre-training is always helping in their low-data regime and emphasize that the chest dataset pre-training yields the best results. However, their miscellaneous body parts radiography dataset only performed comparably to ImageNet, showing the trade-off between data amount and domain gap for choosing a pre-training dataset.

Gonthier et al. studied similar aspects for artwork classification, where pre-training on ImageNet outperformed another artwork pre-training dataset containing around 80k images, presumably ascribed to the low quantity of the artwork pre-training dataset \cite{gonthier2020analysis}. Interestingly, a gradual two-step pre-training, using first ImageNet and then the intermediate artwork dataset, led to further improvement. This suggests that successive pre-trainings could help to bridge domain gaps continuously. 

\subsubsection{Unsupervised domain adaptation}

Unsupervised domain adaptation (UDA) describes training a model with labeled data from a source domain and unlabeled data from a target domain to perform well on the latter. This is of major interest when the target domain labeling process is costly or when source domain annotations are readily available. One very good example is the well-known GTA5 \cite{GTA5} or SYNTHIA \cite{SYNTHIA} (source) to Cityscapes \cite{cityscapes} (target) task. In both cases, source datasets provide synthetic and inherently annotated urban landscapes.

A wide branch of the UDA methods is now improving an adversarial learning framework, which was initially proposed by Ganin et al. \cite{pmlr-v37-ganin15}. The underlying idea is to force the model to adopt a domain-independent feature representation. This is achieved by passing images of both domains into the main model and feeding intermediate layer feature representations into a discriminator. This discriminator then guesses whether the initial data comes from the source or the target domain. Thereby, the discriminator penalizes internal feature representations when they differ substantially for both domains. At the same time, the source data is used in a supervised fashion to train the main model for the task. This adversarial approach has been adapted to semantic segmentation by Tsai et al., where matching of internal features and segmentation masks was performed \cite{tsai2020learning}.

To the largest extent, recent articles about UDA report results on the previously mentioned GTA5 to Cityscape task, and only a few applications can be found in other fields such as the adversarial training approach in the medical domain \cite{Dou2018,zhang2019collaborative}. This techniques' potential for the materials science community has never been showcased to the best of our knowledge. By doing so, we hope that this work will spark interest in UDA techniques in this field. 

\subsection{Contributions}

The contributions of this work are the following:
\begin{itemize}
    \item We study the impact of different data augmentation policies, models, and pre-trainings on the segmentation performance.
    \item We show the impact of these different training strategies on the domain generalization (i.e., applicability) of a model to an alternate target domain. Specifically, the domain generalization across SEM micrographs of differently contrasted complex phase steel microstructures and across different imaging modalities is addressed.
    \item We implement a state-of-the-art UDA approach \cite{tsai2020learning} and show its merit in materials science despite data limitations by bridging the aforementioned domain gaps. The UDA frameworks' suitability for applications beyond the ones presented here is discussed.
\end{itemize}
\section{Methods}

\subsection{Specimen fabrication and image acquisition methodology}
\label{sec:methods}

This work is based on SEM images of a low-carbon complex-phase steel which was subject of our last work \cite{Durmaz2021}. The specimen were taken from thermomechanically rolled heavy industrial plates. Resulting microstructure is composed of lath-shaped bainite (foreground) as well as polygonal and irregular ferrite with dispersed granular carbon-rich 2nd phase and martensite-austenite (MA) islands (collectively referred to as background). Micrographs were taken in the plate’s transversal direction (TD), between quarter- and mid-thickness of the plate. Rolling-induced stress and cooling rate gradients result in a small microstructure variance, where in some images taken from comparatively surface-near regions, the polygonal ferrite grains are elongated in the rolling direction (RD). During imaging, segregation zones in the plate core were avoided.
The specimens were ground using 80–1200 grid SiC papers, and then subjected to polishing with 6, 3, and finally, 1\,µm diamond grain sizes. Using different etching and imaging conditions as well as image modalities, four image sets were drawn from these specimens. The configurations are presented in Table \ref{tab:datasets}.\\

\begin{figure}[htbp]
	\centering
	\footnotesize
	\includegraphics[width=\linewidth]{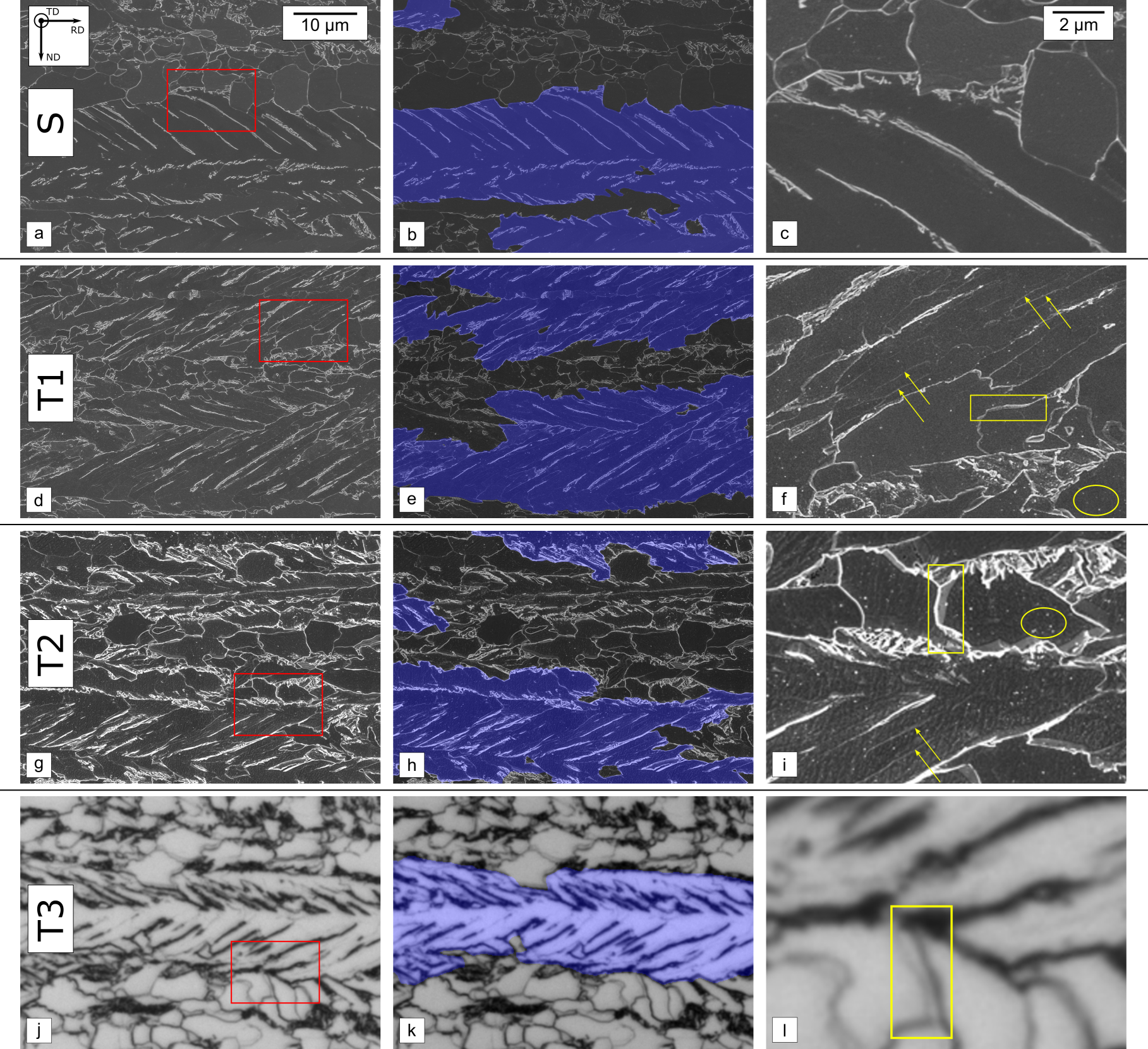}
	\caption{Input images, same superimposed with the annotation, and a detail view are each shown for the source (a--c), target 1 (d--f), target 2 (g--i), and target 3 (j--l) domain. Please refer to Table \ref{tab:datasets} for the technical details about the different sets. The red frames in the first column highlight the region of interest shown in the third column. Yellow annotations are used in Section \ref{sec:Dataset} for discussing major differences between the domains.}
	\label{fig:datasets}
\end{figure}

\noindent \textbf{Etching.} The duration was controlled by a metallographer waiting for a macroscopic contrasting to be visible to the naked eye. Etching reveals grain boundaries since the reaction kinetics depend on the local chemical composition and crystallographic orientation. Therefore, carbide films and a few MA constituents are exposed. An example input image, the same with superimposed label, and a detail view of each data set (i.e., domain) is presented in Figure \ref{fig:datasets}.

\begin{table*}[hbtp]
	\centering
	\caption{Micrograph set descriptions. Mean lath-bainite phase fractions $\phi_{train}$ of the train tiles are given. Note that no annotation was available for the target 1 and target 2 training sets. Hence, their phase fractions were estimated based on pseudo-labels and confirmed by a metallography expert. Symbols "o" and "+" refer to normal and deliberately prolonged etching duration, respectively.} 
	\begin{tabularx}{\linewidth}{
	p{\dimexpr.14\linewidth-2\tabcolsep-1.3333\arrayrulewidth}
    p{\dimexpr.30\linewidth-2\tabcolsep-1.3333\arrayrulewidth}
    p{\dimexpr.17\linewidth-2\tabcolsep-1.3333\arrayrulewidth}
    p{\dimexpr.14\linewidth-2\tabcolsep-1.3333\arrayrulewidth}
    p{\dimexpr.16\linewidth-2\tabcolsep-1.3333\arrayrulewidth}
    p{\dimexpr.08\linewidth-2\tabcolsep-1.3333\arrayrulewidth}
	}
		 \textbf{Set name} & \textbf{Etching} & \textbf{Imaging setup} & \textbf{\# train tiles} & \textbf{\# test images} & \textbf{$\phi_{train}$}\\
		\thickhline
		Source (S) & Struers A2 electrolytic etching; o & SEM setup 1 & 112 & 7 ($\equalhat$ 28 tiles) & 0.53\\ 
		\hline
        Target 1 (T1) & Nital (2 vol. \% HNO$_3$); o & SEM setup 1 & 80 & 5 ($\equalhat$ 20 tiles) & 0.54\\
		\hline
		Target 2 (T2) & Nital (2 vol. \% HNO$_3$); + & SEM setup 2 & 48 & 3 ($\equalhat$ 12 tiles) & 0.29\\
		\hline
		Target 3 (T3) & Nital (2 vol. \% HNO$_3$); + & LOM & 234 & 50 ($\equalhat$ 200 tiles) & 0.53\\
		\thickhline
	\end{tabularx}
	\label{tab:datasets}
\end{table*}

\newpage
\noindent \textbf{Imaging conditions.} The SEM setup 1 was utilizing a Zeiss Merlin FEG-SEM using secondary electron contrast (in-lens) at a magnification of 2000$\times{}$ with an image size of 2048$\times{}$1433 (annotation bar cropped), which represents 56.7$\times{}$42.5\,$\upmu$m$^2$ (pixel size = 27.7\,nm). The SEM was set at an acceleration voltage of 5\,kV, a probe current of 300\,pA, and a working distance of 5\,mm. Small acceleration voltages reduce the interaction volume and increase surface sensitivity.
In SEM setup 2, the micrographs were recorded in a Zeiss Supra FEG-SEM with another in-lens detector. The acceleration voltage, the probe current and the working distance were the same as in the first setup. The magnification was set at 1000$\times{}$ with a higher image resolution, giving virtually identical physical pixel size compared to the first setup. Due to subsequent stitching and cropping, images of SEM setup 2 ultimately had the same size as others. Contrast and brightness settings differed from the first imaging setup, see Figure \ref{fig:datasets}. Within each domain, micrographs were acquired with the same image contrast and brightness settings in the SEM. Lastly, the LOM images were recorded in an Olympus LEXT OLS 4100. Micrographs were taken at a magnification of 1000$\times{}$ with an image resolution of 1024$\times{}$1024 pixels, corresponding to an area of 129.6$\times{}$129.6\,$\upmu$m$^2$ (pixel size = 126.6\,nm). All LOM images were acquired with the same exposure settings.

\noindent \textbf{Annotations.} Segmentation masks were drawn manually by human experts on a digital drawing tablet (Wacom). Correlative EBSD maps on the source domain micrographs rendered the annotation more reproducible and accurate. For more details on the acquisition, multi-modal registration and annotation process we refer to \cite{Durmaz2021}. Note that for the target domains, with the exception of target 3 (T3), annotations were only available for a small portion, i.e. the test images.

\subsection{Dataset processing and descriptions}
\label{sec:Dataset}

\textbf{Processing.} The LOM images were assimilated to the native SEM datasets in terms of physical pixel size and field of view by scaling and cropping operations, cf. Figure \ref{fig:datasets}a and  \ref{fig:datasets}j. Subsequent processing was identical for all datasets. Images were mirror-padded to make them square so four tiles can be extracted per image. Following the results of our last paper \cite{Durmaz2021}, the images were downscaled by a factor $\times{}$0.5 before tiling, effectively increasing the context passed to the models' receptive field\cite{luo2017understanding}. The obtained training tiles are then of size 636$\times{}$636 (512$\times{}$512 plus an additional 62 pixel overlap extending into the adjacent tiles at each tile border). This overlap-tile strategy was introduced by Ronneberger et al. \cite{Ronneberger2015} to cope with memory restrictions and helps the model to circumvent tile border effects. For segmentation loss computation, the overlap regions extending in neighboring tiles were discarded (i.e., 512$\times{}$512 center region was used solely). To align the domains' phase fractions as much as possible, few training tiles of the target domains were discarded, ultimately resulting in the tile numbers and phase fractions listed in Table \ref{tab:datasets}. The source dataset was split in five folds for cross-validation. In contrast, the annotated portion in the target domains was too small to perform cross-validation. Therefore, only a single train and test set has been built for the target datasets. While training was performed with tiles, the model evaluation was conducted on full images. Bright-field LOM images (T3) pixel values were inverted to give them a dark background similar to SEM.

\noindent \textbf{Descriptions.} The microscopic differences between the four datasets become evident in the detail views in Figure \ref{fig:datasets}c, f, i, l. Conducted electrolytical Struers A2 etching (source) does not emphasize the sub-grain boundaries and culminates in comparatively slender carbide film appearance overall. In contrast, target 1 (T1) and to a lesser extent target 2 (T2), highlight the hierarchical structure in the lath-bainite regions (see arrow annotations in Figure \ref{fig:datasets}f, i). However, these two target domains are also accompanied by etching artifacts (see ellipse annotations). In T2 and T3, as highlighted by the rectangle annotations, some grain boundaries appear faded due to their grain boundary inclination and the progressed etching state. Along with the pronounced contrast and wider carbide films, this renders it evident that T2 was over-etched. In the bright-field LOM images (T3), carbide film morphology cannot be resolved and image features differ substantially due to the modality change. In conclusion, these target domain datasets represent typical but distinct degrees of domain gap, where the shift is gradually increasing with respect to the source domain. On the other hand, statistical differences are present particularly in the T2 domain, where the lath-bainite phase fraction deviates from the source significantly.

\subsection{Pre-training datasets}

The first pre-training dataset used in this study is ImageNet\cite{Deng2009ImageNetAL}. Further, the apparent domain gap between ImageNet and our target datasets motivated us to test an additional pre-training dataset with a smaller domain gap. Therefore, we selected a SEM dataset of nanoscientific objects \cite{aversa2018first}, which comprises approximately 22k images, non-uniformly distributed in 10 classes: biological, fibers, films and coated surfaces, microelectromechanical systems (MEMS) and electrodes, nanowires, particles, porous sponges, pattern surfaces, powders, and tips. In the following, we refer to this pre-training dataset with \textit{NanoSEM}.

Before using the NanoSEM dataset, a pre-processing cleaning step was performed. Images with many burned-in measurement annotations were discarded and SEM annotation bars were cropped to avoid spurious correlations between annotation bars and class predictions, known to occur otherwise. Finally, the pre-training dataset amounted to 18,750 images. More details about the pre-training methodology and results are supplied in Section \ref{subsec:PT} and the Supplemental.

\subsection{Deep learning training methodology}
\subsubsection{Segmentation architectures}

As part of this work, two main segmentation architectures are implemented. The well-established U-Net \cite{Ronneberger2015} is used in the first place to investigate different pre-training strategies. This fully-convolutional architecture is, as its name implies, composed of an encoder-decoder structure with skip connections between the corresponding levels of the encoder and decoder. It gave outstanding results on medical segmentation tasks even with very little training data.

Among many different segmentation models that were proposed after the U-Net, one series of models marked a turning point in this field. Chen et al. published the first version under the name DeepLab \cite{chen2017deeplab}. In this work, the second DeepLab version is implemented (DeepLabv2). This architecture uses so-called dilated convolutions (or atrous convolutions), which help the model to enlarge its field of view (receptive field) and take patterns at larger scales into account appropriately. The main idea of DeepLabv2 is to learn and aggregate patterns at different scales with dilated convolutions having different dilation rates. This aggregation of dilated convolutions effectively causes a more uniform distribution in the effective receptive field \cite{luo2017understanding}.

The encoder used for the U-Net is a portion of the VGG16 classification network \cite{simonyan2015deep}, while DeepLabv2 was built with a ResNet-101 \cite{he2015deep} encoder. The exact architecture for both cases are given in the Supplemental and in \cite{tsai2020learning}, respectively.

For segmentation training, a binary cross-entropy loss and an Adam optimizer was employed. Learning rates, batch sizes and training times vary along this study. Thus, these parameters will be specified in Section \ref{sec:results} for the different experiments. The models were trained on a GPU cluster node consisting of four parallel NVIDIA Ampere A100 GPUs.

\subsubsection{Pre-training \& fine-tuning procedure for domain generalization experiments}
\label{subsec:PT}

For the ImageNet pre-training, we used pre-trained weights provided by the python package \textit{Segmentation models pytorch} \cite{Yakubovskiy:2019}.

Concerning our self-performed NanoSEM pre-training, we passed the U-Net encoder output to an auxiliary classification head \cite{Yakubovskiy:2019}. This head consists of a global average pooling layer, followed by 50\% dropout and a linear layer with a sigmoid activation. The auxiliary classification head facilitates encoder training on classification datasets.

\begin{figure}[!htbp]
	\centering
	\footnotesize
	\includegraphics[width=\linewidth]{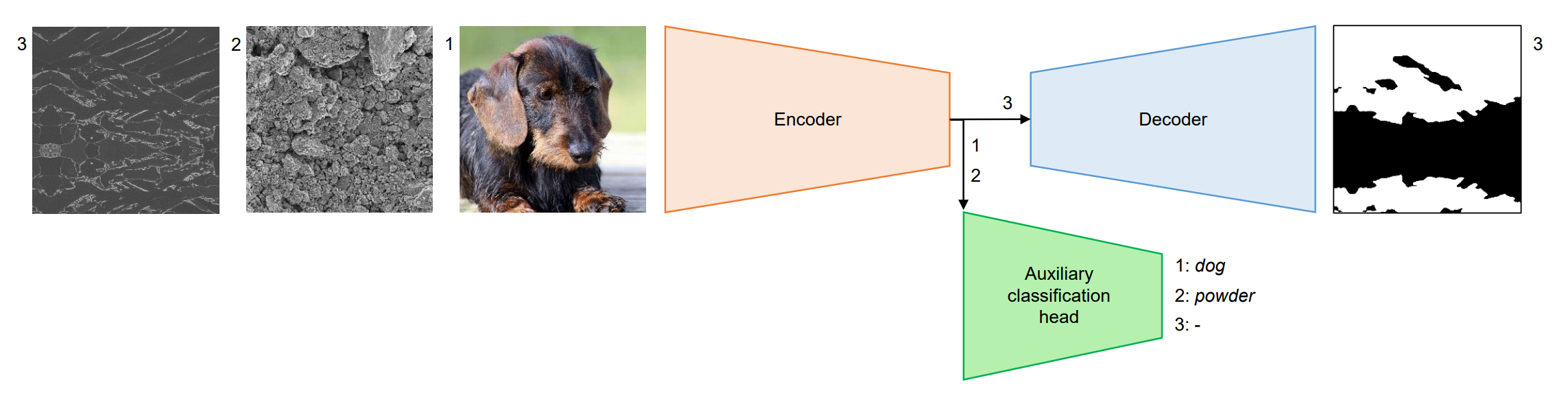}
	\caption{Summary of the pre-training and fine-tuning procedure. Starting from ImageNet pre-trained weights (1), we optionally use the NanoSEM dataset for pre-training the encoder of the model on this classification task (2). The weights of the encoder are then directly transfered for the final fine-tuning on the source domain (3), while the decoder is random-initialized.}
	\label{fig:PT_process}
\end{figure}

For NanoSEM pre-training, ImageNet weights were used as an initialization for the trainings, making this process a two-step pre-training (from ImageNet to NanoSEM to the final task). The 18,750 NanoSEM images were split into 15k images for training and 3,750 images for testing purposes. Pre-training used an Adam optimizer with a constant and encoder-layer independent learning rate and was run for 100 epochs. The obtained pre-trained models were transferred to the segmentation task by just copying the weights of the model performing best on the pre-training task. The full model was then fine-tuned on the source dataset (without frozen layers) with a reduced learning rate for the pre-trained encoder (10$\times{}$ lower than the decoder learning rate). The two-stage pre-training process along with fine-tuning is summarized in Figure \ref{fig:PT_process}. Please note that the individual learning rates applied at the pre-training and fine-tuning stage are of major importance. An optimization of the learning rate used for pre-training on NanoSEM has been carried out and is given in the Supplemental. In case of the VGG16 U-Net model, aside from random initialization, either the two-stage pre-training or ImageNet pre-trained weights were used as initial conditions before pursuing fine-tuning to the source domain. This allowed to investigate the impact of pre-training on generalization capability to the target domains (see Table \ref{tab:datasets}). In contrast, for the DeepLabv2 models solely ImageNet pre-trained weights were used. The described procedures for both architectures, did not utilize any target domain data. Instead, source domain fine-tuned models were directly tested on target domain data (domain generalization).

\subsubsection{Unsupervised domain adaptation --- An adversarial framework}
\begin{figure}[htbp]
	\centering
	\footnotesize
	\includegraphics[width=\linewidth]{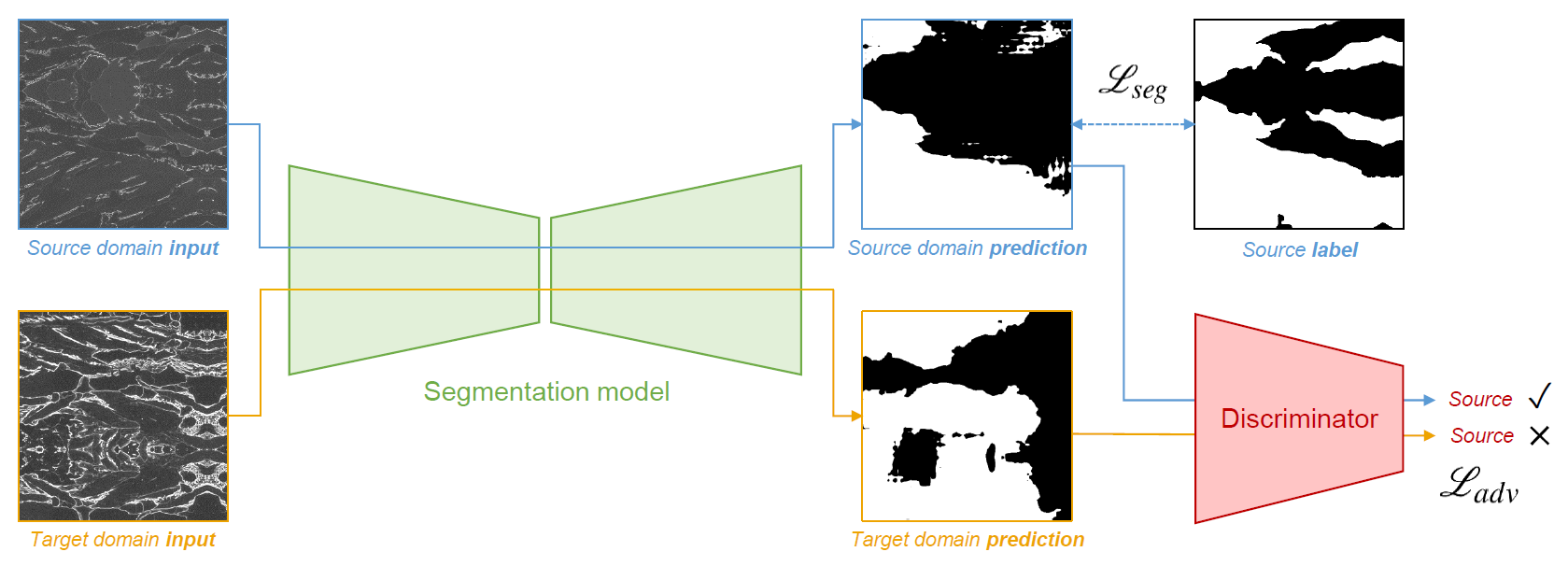}
	\caption{Simplified schematic representation of AdaptSegNet \cite{tsai2020learning}. Tile examples from the source and target 2 domains are given on the left. The source prediction is used in a supervised manner with cross-entropy loss. Both domains' prediction are fed into the discriminator.}
	\label{fig:UDA_model}
\end{figure}

To additionally take advantage of \textit{unlabeled} target domain data, which often can be created in abundance effortlessly, the UDA method of Tsai et al.\cite{tsai2020learning} has been implemented.
This adversarial framework can be used to train a semantic segmentation unsupervised domain adaptation task. It is based on the original idea proposed by Ganin et al. \cite{pmlr-v37-ganin15}
The code of\cite{tsai2020learning} was adapted to make it compatible with our data. Figure \ref{fig:UDA_model} depicts the training process in a simplified fashion.

As proposed in the original framework \cite{tsai2020learning}, we utilize a DeepLabv2 with a ResNet-101 as the segmentation architecture, which facilitates comparability with the corresponding domain generalization experiments described at the end of Section \ref{subsec:PT}. In the UDA framework, annotated source and unannotated target domain data are fed into this segmentation model (shared weights). The source domain prediction is used for training the segmentation model in a supervised manner, given that labels are available in this domain. This gives a first part of the loss function ($\mathcal{L}_{seg}$), evaluated as a binary cross-entropy. Furthermore, source and target domain predictions are passed to a discriminator model, which attempts to classify from which domain the prediction comes. The second part of the loss, the so-called adversarial loss ($\mathcal{L}_{adv}$), quantifies the ability of the segmentation model to fool the discriminator. It is also computed as a binary cross-entropy for the domain classification. Additional to the segmentation outputs, network-internal feature representations of both domains are extracted from an auxiliary segmentation head, reshaped to segmentation mask size (auxiliary segmentation), and passed to the discriminator ($\mathcal{L}^{aux}_{adv}$). This is not represented on Figure \ref{fig:UDA_model} for the sake of simplicity. Moreover, the source domain auxiliary segmentation is compared to the annotation mask ($\mathcal{L}^{aux}_{seg}$). The different loss parts are weighted so emphasize can be put on either of the segmentation or adversarial losses introducing three further hyperparameters. Note that the loss portions related to the auxiliary feature output are typically less weighted, making Figure \ref{fig:UDA_model} representation a good first approximation of the model. The four aforementioned loss parts ($\mathcal{L}_{seg},\mathcal{L}_{adv},\mathcal{L}^{aux}_{adv},\mathcal{L}^{aux}_{seg}$) compose the training loss of the segmentation model, whereas the discriminator is optimized based on a domain-classification cross-entropy loss. When back-propagating the combined loss of the segmentation model, the weights of the discriminator are temporarily frozen.
Both the segmentation and discriminator models are trained in an end-to-end fashion.
Full technical details are given in the Supplemental. 

This complex architecture forces the segmentation model to learn features that are domain-independent, rendering the transfer from the source to the target domain possible. 

\subsubsection{Data augmentation}

Data augmentation is a common practice in ML for increasing the labeled data amount without additional annotation cost. It consists in applying transformations to the data before passing it to the model. The main objective is to render the network invariant to specific transformations.
As part of this work, a simple flip and 90° rotation pipeline (with probability 0.5) was first tested (marked as basic). Moreover, an extended pipeline making use of, amongst others, elastic transformation and optimized for the source domain in our last publication \cite{Durmaz2021} was implemented (marked as extended). Both pipelines have been built with the Albumentations package \cite{Albumentation}. Full details about the pipelines are given in the Supplemental. Data augmentation was applied to train all our models except for the NanoSEM pre-training. In the UDA framework, both source and target datasets were augmented using the optimized pipeline.

\subsubsection{Evaluation metrics}

For the segmentation models evaluation, we use an intersection over union (IoU) metric, see Equation \ref{eq:IoU}, averaged over background and foreground classes (mIoU). In order to evaluate the trained models' generalizability, we used a relative mIoU deviation between the source domain performance with sole supervised learning and the target domain performance with the concerned generalization method (either domain generalization or UDA). We refer to this metric as \textit{relative domain transferability} (RDT; see Equation \ref{eq:RDT}). Using a \textit{relative} deviation avoids overrating models that are performing better on the target domains because of their inherent advantage on the source domain. For instance, a hypothetical model with 70\% and 65\% mIoU on the source and target (RDT = -0.07) generalizes better than one yielding 80\% and 70\% mIoU (RDT = -0.13), despite latters' better target performance.

\begin{gather}
    IoU = \frac{TP}{TP+FP+FN}. \label{eq:IoU}\\
    RDT = \frac{mIoU_{T}-mIoU_{S}}{mIoU_{S}}. \label{eq:RDT}
\end{gather}

$TP$, $TN$, $FP$, and $FN$ represent true positive, true negative, false positive, and false negative pixels, respectively. $mIoU_{T}$ and $mIoU_{S}$ are the class-averaged model performance on the target domain and reference source domain, respectively.

\section{Results}
\label{sec:results}

In the first part (Section \ref{sec:sup_learn_on_source}), the fully supervised results on the source domain will be presented, posing a reference for the segmentation task. Subsequently, performance on the target domains (T1,T2,T3, see Table \ref{tab:datasets}) is tested in Section \ref{sec:res_domain_generalization} by directly applying former models (domain generalization = DG) and by unsupervised domain adaptation (UDA), showing the latter methods' benefit.

\subsection{Supervised-learning on the source domain}
\label{sec:sup_learn_on_source}

Models were trained on the source dataset in a supervised fashion. Five-fold cross-validation was implemented to investigate the influence of different training data configurations. The VGG16 U-Net architecture has been used to test aforementioned pre-training and data augmentation settings in order to maximize the phase segmentation task performance. A batch size of 12 with a constant learning rate of $\lambda = 5E-3$ was utilized. The models were trained for 200 epochs, except for experiment S.3, which required more iterations due to the random initialization coupled with the extended augmentation pipeline (increased data variance). This experiment was thus extended to 400 epochs.  

Additionally, a ResNet-101 DeepLabv2 model was trained in a supervised manner. This segmentation architecture corresponds to the one used in the UDA framework, rendering the results comparable. For this architecture, only the pre-training on ImageNet was implemented. The models were trained for 550 epochs with an initial learning rate of $\lambda = 1E-3$ and a polynomial decay with a decay factor of 0.9. Note that a larger epoch number was required for this architecture because of ResNet-101's higher number of model parameters. The mentioned training epoch numbers were chosen to ensure proper convergence of the validation mIoU. The results are given in Table \ref{tab:source}.

\begin{table}[ht]
\caption{Performance of the models trained on the source dataset (cf. Table \ref{tab:datasets}). The given mIoU is always averaged over the five folds used for cross-validation. NanoSEM refers to the two-step pre-training introduced in Section \ref{subsec:PT}. The augmentation pipelines are detailed in the Supplemental.}

	\centering
	\begin{tabularx}{\linewidth}{
	p{\dimexpr.35\linewidth-2\tabcolsep-1.3333\arrayrulewidth}
    p{\dimexpr.18\linewidth-2\tabcolsep-1.3333\arrayrulewidth}
    p{\dimexpr.19\linewidth-2\tabcolsep-1.3333\arrayrulewidth}
    p{\dimexpr.12\linewidth-2\tabcolsep-1.3333\arrayrulewidth}
    p{\dimexpr.16\linewidth-2\tabcolsep-1.3333\arrayrulewidth}
	}
	 \textbf{Model} & \textbf{Pre-training} & \textbf{Augmentation} & \textbf{Exp.\#} &
	 \textbf{mIoU [\%]}\\
	\thickhline
    \multirow{9}{*}{VGG16 U-Net} & \multirow{3}{*}{Random init.} & --- & S.1 & 75.8 ± 2.4\\
    \cline{3-5} & & basic & S.2 & 77.8 ± 2.0\\
    \cline{3-5} & & extended & S.3 & 78.7 ± 2.6\\       
    \cline{2-5} & \multirow{3}{*}{ImageNet} & --- & S.4 & 76.1 ± 3.0\\
    \cline{3-5} & & basic & S.5 & 78.9 ± 2.2\\
    \cline{3-5} & & extended & S.6 & 79.2 ± 2.2\\
    \cline{2-5} & \multirow{3}{*}{NanoSEM} & --- & S.7 & 76.9 ± 2.6\\
    \cline{3-5} & & basic & S.8 & 79.2 ± 2.2\\
    \cline{3-5} & & extended & S.9 & \textbf{80.2} ± 2.4\\   
    \hline
    ResNet-101 DeepLabv2 & ImageNet & extended & S.10 & 79.4 ± 1.8\\ 
	\thickhline
\end{tabularx}
\label{tab:source}
\end{table}

The results in Table \ref{tab:source} demonstrate that pre-training helps the model in this low-data regime. Indeed, we observe the systematic trend that the two-stage NanoSEM pre-training outperforms the ImageNet one, which in turn surpasses random initialization, regardless of the used augmentation pipeline. Similarly, the extended augmentation pipeline consistently grants better results than the basic one, which in turn compares favorably to the unaugmented result.
Thus, the best performance is observed with the NanoSEM pre-training and the extended data augmentation pipeline, reaching 80.2\% mIoU as an average over the five folds (S.9).

Experiment S.10 consists in the supervised training of the DeepLabv2 architecture on the source domain. Concerning the ImageNet pre-training case, DeepLabv2 slightly outperforms the VGG16 architecture on this task (cf. S.10 to S.6). 

In the following sections, all mIoU values in Table \ref{tab:source} are utilized as the reference value to compute equivalent DG experiments' RDT metrics. Analogously, model S.10 serves as the reference for all the UDA-based models. 

\subsection{Model generalization and domain adaptation to target domains} \label{sec:res_domain_generalization}

In this section, the main objective is to achieve good segmentation results on the target datasets introduced in Table \ref{tab:datasets} even though no labeled training data is available in these domains. The three following subsections are dedicated to the three target datasets with continuously increasing domain shifts.

\subsubsection{Etching type variation --- A small domain shift to start with (T1)}

\begin{figure}[!htbp]
	\centering
	\footnotesize
	\includegraphics[width=0.72\linewidth]{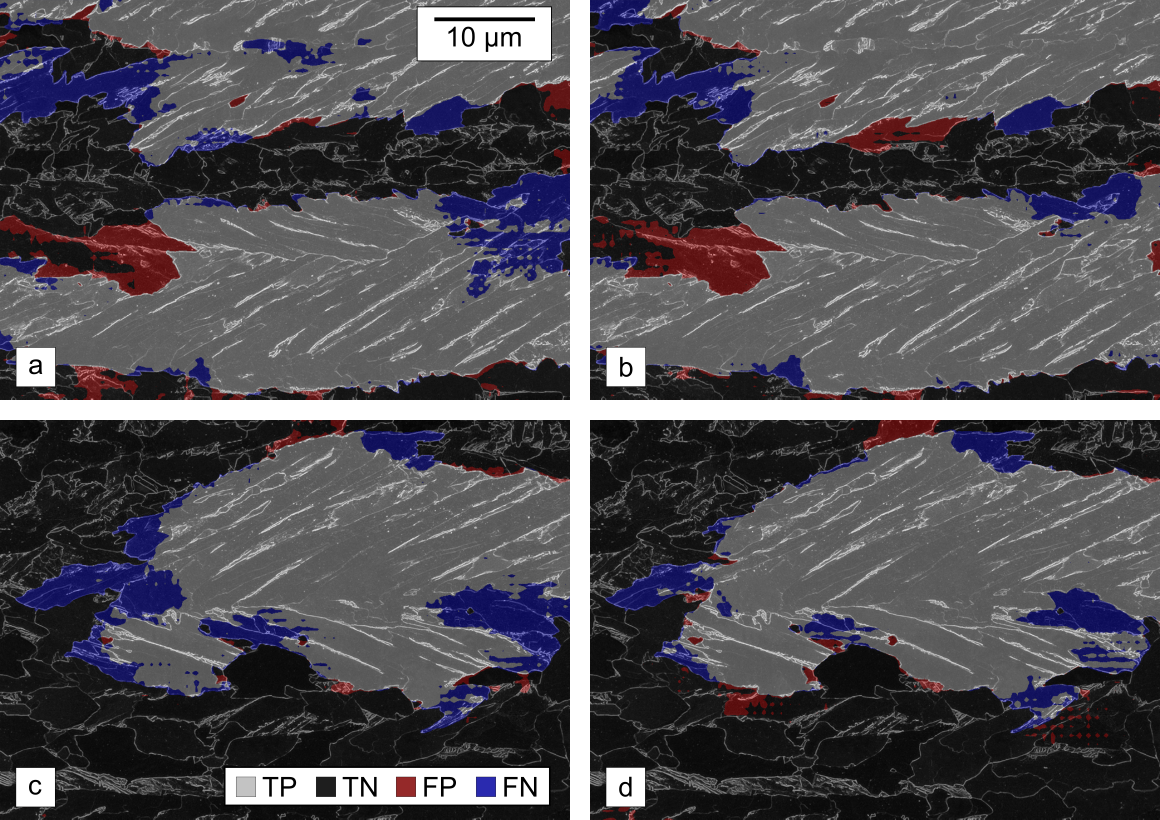}
	\caption{Predictions on the target 1 test images. The left column (a),(c) gives the predictions of a DeepLabv2 model trained on the source dataset (cf. experiment T1.10). The right column (b),(d) gives the predictions of a DeepLabv2 model trained with the UDA framework on the source and target 1 datasets (cf. experiment T1.11).}
	\label{fig:before-after_Target1}
\end{figure}

The models trained in Section \ref{sec:sup_learn_on_source} were tested on this target domain, and the aforementioned UDA framework was used for improving the results. These UDA models were trained with a polynomial learning rate ($\lambda = 1E-3$, decay factor of 0.9) for 3,000 epochs and a batch size of 8 both for source and target tiles. Once again, the number of epochs has been chosen to reach a satisfying convergence of the validation mIoU. The results are given in Table \ref{tab:target1} and some visualizations are provided in Figure \ref{fig:before-after_Target1} to show the advantage of the UDA method over domain generalization.

\begin{table}[ht]
\caption{Performance of the models trained on the source domain evaluated on the target 1 domain (cf. Table \ref{tab:datasets}) along with the UDA reached performance. The given mIoU is always averaged over the five folds used for cross-validation. NanoSEM refers to the two-step pre-training introduced in Section \ref{subsec:PT}. The augmentation pipelines are detailed in the Supplemental.}

	\centering
	\begin{tabularx}{\linewidth}{
	p{\dimexpr.3\linewidth-2\tabcolsep-1.3333\arrayrulewidth}
	p{\dimexpr.17\linewidth-2\tabcolsep-1.3333\arrayrulewidth}
    p{\dimexpr.17\linewidth-2\tabcolsep-1.3333\arrayrulewidth}
    p{\dimexpr.11\linewidth-2\tabcolsep-1.3333\arrayrulewidth}
    p{\dimexpr.14\linewidth-2\tabcolsep-1.3333\arrayrulewidth}
    p{\dimexpr.11\linewidth-2\tabcolsep-1.3333\arrayrulewidth}
	}
	\textbf{Model} & \textbf{Pre-training} & \textbf{Augmentation} & \textbf{Exp.\#} &
	\textbf{mIoU [\%]} & \textbf{RDT [\%]}\\
	\thickhline
    \multirow{9}{*}{\textbf{DG} - VGG16 U-Net} & \multirow{3}{*}{Random init.} & --- & T1.1 & 75.7 ± 2.2 & -0.1\\
    \cline{3-6} 
    & & basic & T1.2 & 77.5 ± 1.4 & -0.4\\
    \cline{3-6} 
    & & extended & T1.3 & 73.5 ± 3.9 & -4.2\\       
    \cline{2-6} 
    & \multirow{3}{*}{ImageNet} & --- & T1.4 & 77.2 ± 1.0 & \;1.4\\
    \cline{3-6} 
    & & basic & T1.5 & 79.8 ± 1.1 & \;1.2\\
    \cline{3-6} 
    & & extended & T1.6 & 81.6 ± 1.9 & \;3.1\\
    \cline{2-6} 
    & \multirow{3}{*}{NanoSEM} & --- & T1.7 & 77.2 ± 0.6 & \;0.4\\
    \cline{3-6} 
    & & basic & T1.8 & 80.6 ± 0.8 & \;1.8\\
    \cline{3-6} 
    & & extended & T1.9 & 81.5 ± 0.7 & \;1.6\\   
    \hline
    \textbf{DG} - ResNet-101 DeepLabv2 & ImageNet & extended & T1.10 & 82.2 ± 1.5 & \;3.6\\ 
	\thickhline
	\textbf{UDA} - ResNet-101 DeepLabv2 & ImageNet & extended & T1.11 & 84.7 ± 0.7 & \;6.7\\ 
	\thickhline
\end{tabularx}
\label{tab:target1}
\end{table}

Looking at the results from Table \ref{tab:target1}, it first appears surprising that most source domain trained models perform excellent on T1, sometimes even exceeding the source performances, resulting in positive RDT values. This will be discussed in Section \ref{sec:generalization_of_trained_models}. Secondly, the general tendency that pre-training helps domain generalizability is evident, as random-initialized models yield the lowest RDT values.
Despite the aforementioned good model transferability between source and T1, UDA surpasses DG clearly (compare T1.11 to T1.10). Aside from the 2.5\% mIoU increase in favor of the UDA approach, the obtained model is more balanced in terms of class mispredictions. Figure \ref{fig:before-after_Target1} illustrates this phenomenon with two examples. Without UDA, the models transferred from scratch exhibit a skewed behavior towards the background class, thus giving substantially more false negatives than false positives. On Figure \ref{fig:before-after_Target1}b, d, the amount of false negatives is reduced while false positives increase slightly, giving an overall better segmentation and an improved phase fraction estimation in the UDA case.
Additionally, it can be observed on Figure \ref{fig:before-after_Target1} that both models (DG and UDA) detect the lath-shaped areas as bainite appropriately, whereas the DG model particularly struggles to find the proper boundaries between the foreground and background classes. The main improvement offered by UDA lies in the better boundary localization (Figure \ref{fig:before-after_Target1}b, d).
\newpage

\subsubsection{Etching type and imaging setup variation --- A larger domain shift to address (T2)}

\begin{figure}[htbp!]
	\centering
	\footnotesize
	\includegraphics[width=0.72\linewidth]{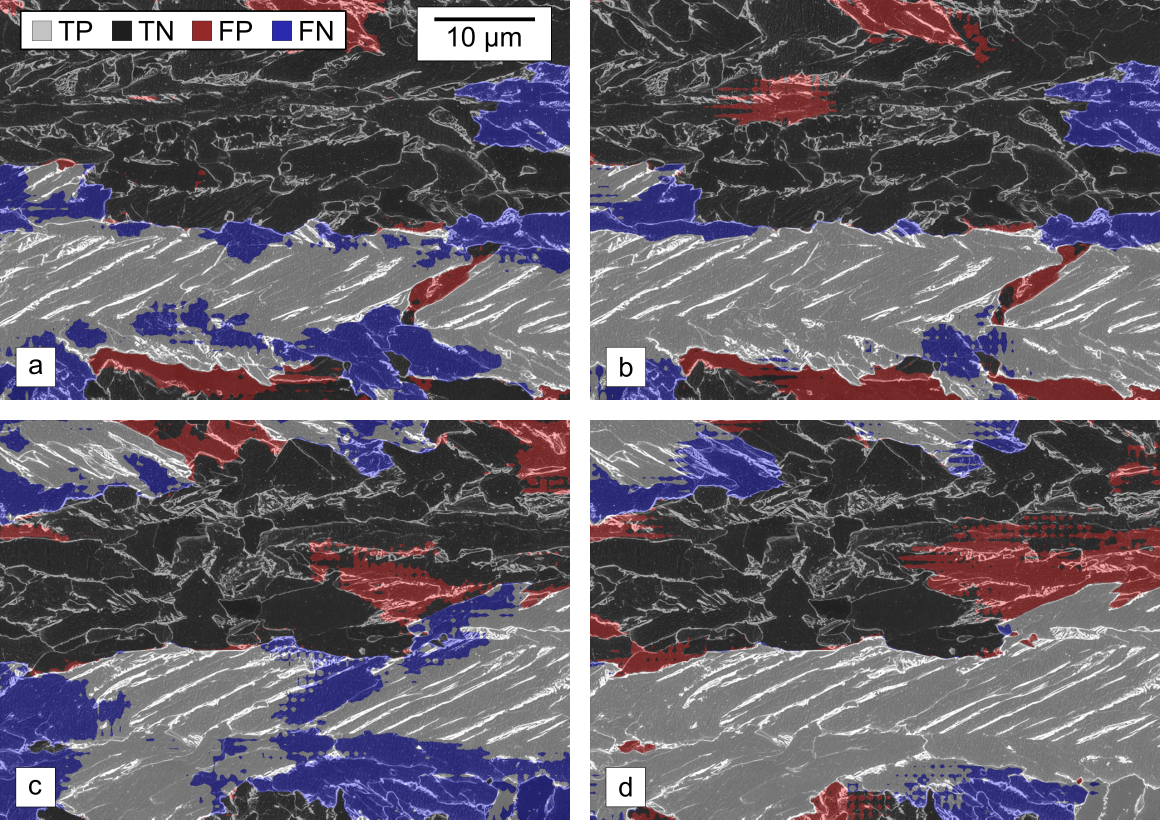}
	\caption{Predictions on the target 2 test images. The left column (a),(c) gives the predictions of a DeepLabv2 model trained on the source dataset (cf. experiment T2.10). The right column (b),(d) gives the predictions of a DeepLabv2 model trained with the UDA framework on the source and target 2 datasets (cf. experiment T2.11).}
	\label{fig:before-after_Target2}
\end{figure}

Analogous to the previous subsection, this one addresses the T2 dataset (cf. Table \ref{tab:datasets}). Results are given in Table \ref{tab:target2} and visualizations on Figure \ref{fig:before-after_Target2}.

\begin{table}[ht]
\caption{Performance of the models trained on the source domain evaluated on the target 2 domain (cf. Table \ref{tab:datasets}) along with the UDA reached performance. The given mIoU is always averaged over the five folds used for cross-validation. NanoSEM refers to the two-step pre-training introduced in Section \ref{subsec:PT}. The augmentation pipelines are detailed in the Supplemental.}

	\centering
		\begin{tabularx}{\linewidth}{
	p{\dimexpr.3\linewidth-2\tabcolsep-1.3333\arrayrulewidth}
	p{\dimexpr.17\linewidth-2\tabcolsep-1.3333\arrayrulewidth}
    p{\dimexpr.17\linewidth-2\tabcolsep-1.3333\arrayrulewidth}
    p{\dimexpr.11\linewidth-2\tabcolsep-1.3333\arrayrulewidth}
    p{\dimexpr.14\linewidth-2\tabcolsep-1.3333\arrayrulewidth}
    p{\dimexpr.11\linewidth-2\tabcolsep-1.3333\arrayrulewidth}
	}
	 \textbf{Model} & \textbf{Pre-training} & \textbf{Augmentation} & \textbf{Exp.\#} &
	 \textbf{mIoU [\%]} & \textbf{RDT [\%]}\\
	\thickhline
    \multirow{9}{*}{\textbf{DG} - VGG16 U-Net} & \multirow{3}{*}{Random init.} & --- & T2.1 & 51.4 ± 4.9 & -32.1\\
    \cline{3-6} & & basic & T2.2 & 57.6 ± 4.9 & -26.0\\
    \cline{3-6} & & extended & T2.3 & 52.5 ± 8.5 & -31.5\\       
    \cline{2-6} & \multirow{3}{*}{ImageNet} & --- & T2.4 & 58.1 ± 2.4 & -23.7\\
    \cline{3-6} & & basic & T2.5 & 63.8 ± 1.1 & -19.2\\
    \cline{3-6} & & extended & T2.6 & 52.8 ± 11.8 & -33.2\\
    \cline{2-6} & \multirow{3}{*}{NanoSEM} & --- & T2.7 & 59.7 ± 3.0 & -22.3\\
    \cline{3-6} & & basic & T2.8 & 61.2 ± 3.7 & -22.7\\
    \cline{3-6} & & extended & T2.9 & 60.3 ± 3.3 & -24.8\\   
    \hline
    \textbf{DG} - ResNet-101 DeepLabv2 & ImageNet & extended & T2.10 & 61.0 ± 3.7 & -23.2\\ 
	\thickhline
	\textbf{UDA} - ResNet-101 DeepLabv2 & ImageNet & extended & T2.11 & 67.3 ± 2.0 & -15.3\\ 
	\thickhline
\end{tabularx}
\label{tab:target2}
\end{table}

Regarding domain generalization (DG), two major observations can be made. First, exactly as for T1, it appears that pre-training helps. Second, as opposed to T1, the basic augmentation pipeline improved over the extended one. Moreover, there is a significant RDT drop in T2, rendering this a more challenging task for the UDA framework.

In this case, UDA gives a pronounced advantage, exceeding the DG DeepLabv2 model by 6.3\% mIoU. Figure \ref{fig:before-after_Target2} displays how UDA corrects the skew towards the background class, leading to a segmentation that is better and more balanced in terms of misclassifications. While difficult regions at the top of Figure \ref{fig:before-after_Target2}a, b remain challenging for the UDA network, larger lath-shaped regions are segmented more comprehensively (cf. bottom of Figures \ref{fig:before-after_Target2}a and \ref{fig:before-after_Target2}b or \ref{fig:before-after_Target2}c and \ref{fig:before-after_Target2}d). The classification of these difficult regions at the top of Figure \ref{fig:before-after_Target2}a, b is equally complicated for humans experts.

Furthermore, some checkerboard patterns appear clearly on Figures \ref{fig:before-after_Target2}c and \ref{fig:before-after_Target2}d. These periodic patterns originate from bilinear interpolation in the DeepLabv2 architecture used to restore the input image resolution after the encoding stage and in regions of model uncertainty. In such uncertain areas the segmentation could be improved by combining models in a voting scheme (i.e.,  bagging) for getting better final predictions. Such a bagging strategy will be briefly discussed in Section \ref{sec:generalization_of_trained_models}.

\subsubsection{Bridging the gap between modalities (T3)}

\begin{table}[ht]
\caption{Performance of the DeepLabv2 model trained on the source domain evaluated on the target 3 domain (cf. Table \ref{tab:datasets}) along with the UDA reached performance. The given mIoU is always averaged over the five folds used for cross-validation.}

	\centering
	\begin{tabularx}{\linewidth}{
	p{\dimexpr.3\linewidth-2\tabcolsep-1.3333\arrayrulewidth}
	p{\dimexpr.17\linewidth-2\tabcolsep-1.3333\arrayrulewidth}
    p{\dimexpr.17\linewidth-2\tabcolsep-1.3333\arrayrulewidth}
    p{\dimexpr.11\linewidth-2\tabcolsep-1.3333\arrayrulewidth}
    p{\dimexpr.14\linewidth-2\tabcolsep-1.3333\arrayrulewidth}
    p{\dimexpr.11\linewidth-2\tabcolsep-1.3333\arrayrulewidth}
	}
	\textbf{Model} & \textbf{Pre-training} & \textbf{Augmentation} & \textbf{Exp.\#} &
	\textbf{mIoU [\%]} & \textbf{RDT [\%]}\\
	\thickhline
    \textbf{DG} - ResNet-101 DeepLabv2 & ImageNet & extended & T3.1 & 49.7 ± 7.6 & -37.4\\ 
	\thickhline
	\textbf{UDA} - ResNet-101 DeepLabv2 & ImageNet & extended & T3.2 & 73.3 ± 0.7 & -7.7\\ 
	\thickhline
\end{tabularx}
\label{tab:target3}
\end{table}

Lastly, the T3 (LOM) dataset has been investigated. Along with the UDA performance, the domain generalization DeepLabv2 model is reported as the sole baseline in Table \ref{tab:target3}. The VGG16 U-Net models are omitted, considering their poor performance on this target domain. In this case, the large scatter between the five folds' results made any conclusion about the relative domain generalizability of individual models on T3 impossible to draw.

\begin{figure}[htbp!]
	\centering
	\footnotesize
	\includegraphics[width=0.72\linewidth]{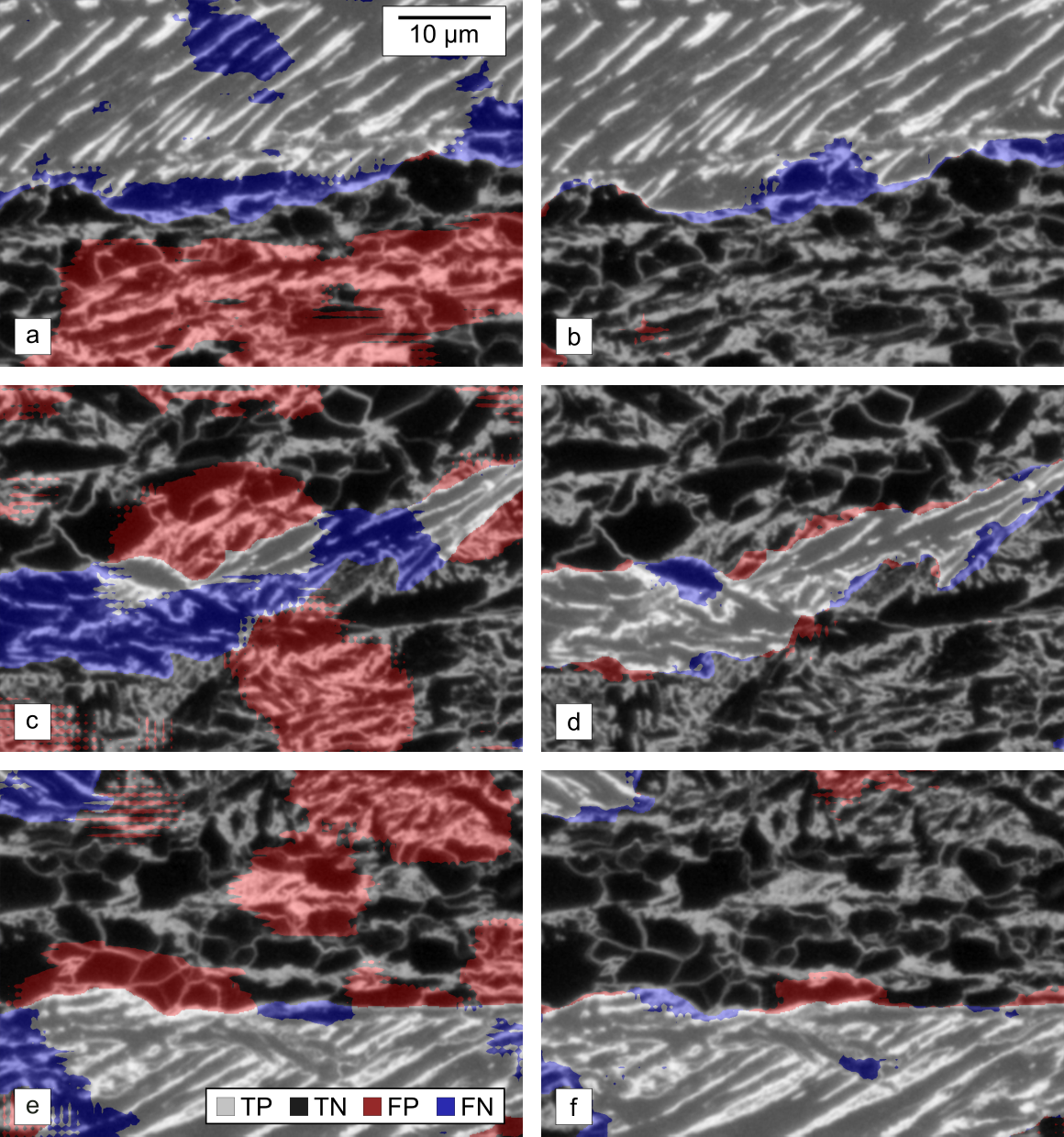}
	\caption{Predictions on the inverted target 3 test images. The left column (a),(c),(e) gives the predictions of a DeepLabv2 model trained on the source dataset (cf. experiment T3.1). The right column (b),(d),(f) gives the predictions of a DeepLabv2 model trained with the UDA framework on the source and target 3 datasets (cf. experiment T3.2).}
	\label{fig:before-after_Target3}
\end{figure}

While domain generalization seems compromised on this target dataset, a tremendous improvement of 23.6\% mIoU is experienced when using the UDA method. Moreover, the scatter over the five folds is reduced substantially. Considering the few prediction examples in Figure \ref{fig:before-after_Target3}, it is apparent that UDA turns a completely unusable model into a convincing one without requiring any labeled data in the target domain.

\newpage
\section{Discussion}
\label{sec:discussion}

\subsection{Supervised-learning on the source dataset}

Achieving the results above with this source data set (see Table \ref{tab:source}) contrasts the common preconception that DL techniques require a large training data quantity. Specifically, satisfactory results on this complex microstructure segmentation task were attained despite using barely more than 100 tiles (27 native SEM images) for training. This can be explained by two factors. First, the data has been acquired in a very repeatable manner. Images of each dataset (i.e., domain) were drawn from an individual etched specimen, and reproducible imaging conditions were applied among images of the same domain. It has to be underlined that this results in comparatively low intra-domain variance, which might not be representative for large-scale datasets acquired by multiple operators or even different institutions. Second, the native micrographs exhibiting a high resolution and rich feature density, 27 such images still represent an appropriate learning foundation for our binary segmentation task.

The results in Table \ref{tab:source} emphasize that pre-training improves the performance of the trained models, giving up to 1.5\% mIoU improvement between the best random-initialized model (S3) and the best NanoSEM pre-trained model (S9). Moreover, pre-trained weight initialization led to faster model convergence. Even with further training, it was observed that random initialized models did not catch up to the pre-trained ones. Therefore, the dataset is situated in the low-data regime mentioned by He et al. \cite{He2019RethinkingIP}, where pre-training elevates the performance irrespective of training iterations. In addition, the two-step NanoSEM pre-training shows better performance compared to ImageNet in all cases.
In contrast, pre-training \textit{solely} on NanoSEM resulted in poor model performance (not reported here).
These observations are in line with Gonthier et al. \cite{gonthier2020analysis}, who performed a two-step pre-training process as well to gradually bridge the domain gap between real-word image datasets and artwork datasets. Please note that this gradual pre-training procedure, compared to conventional pre-training, introduces a further learning rate hyperparameter, which is known to affect the final task performance sensitively. Therefore, relatively more learning rate optimization is required for the pre-training and fine-tuning steps. More details on the learning rate variation are given in the Supplemental.

This two-stage pre-training on ImageNet and NanoSEM might be called into question considering the slight performance increase over sole ImageNet pre-training (1\% improvement from experiment S6 to S9). However, it has to be emphasized that NanoSEM is far from being the optimal pre-training dataset for our target task. Indeed, it entails the following limitations:
\begin{itemize}
    \item Structures in certain classes such as MEMS, patterned surfaces, and tips contain shape-related features but barely any apparent microstructural ones, making the learned weights possibly sub-optimal for the final task.
    \item The image formation is complicated in SEM and depends on a multitude of settings. Most images in the NanoSEM dataset, depending on the class, were either acquired with an Everhart-Thornley (SE2) detector or in-lens detector. Therefore, concerning the detector class, only the latter portion of the pre-training data matches the acquisition of both source and target datasets. Generally, the SE2 detector exhibits a more pronounced topography sensitivity due to its location and orientation, while the in-lens detector combines surface topography and, to a lesser extent, material contrast. 
    \item NanoSEM represents a classification task. Hence, only the encoder of our segmentation model could be pre-trained. While it can be assumed to be domain gap dependent, there is no quantitative understanding to which extent and how many layers of a segmentation model would benefit from such a decoder pre-training. Conventional pre-training was reported to primarily help the models' first layers to learn general features \cite{Yosinski2014HowTA}.
    \item For pre-training standards, NanoSEM is comparatively small.
\end{itemize}

Despite these inadequacies, the underlying rationale of utilizing this NanoSEM dataset for pre-training was that high-level characteristics such as noise levels typical image textures can be learned. However, we assume that a more extensive dataset involving a micrograph segmentation task of arbitrary alloy would prove beneficial over NanoSEM. For instance, The ultra-high carbon steel micrographs collection subset introduced in \cite{decost_lei_francis_holm_2019} would have been appropriate if not for its low quantity. A more promising candidate could be the recently published Aachen-Heerlen annotated steel microstructure dataset\cite{iren2021aachen} containing annotated martensite-austenite islands. While this datasets' annotations exhibit a systematic offset at instance boundaries potentially causing adverse effects during learning, such tendencies presumably can be unlearned during fine-tuning.

With respect to data augmentation, a systematic increase in performance is observed when applying the two pipelines, which is not surprising considering the low amount of data used for training the models.

Lastly, it appears that the DeepLabv2 architecture achieves better results compared to the U-Net one (compare experiments S.6 and S.10). However, the improvement is relatively small considering the model size difference. A possible explanation is that our segmentation task does not exploit the full representation power of the ResNet-101 DeepLabv2 architecture. Another potential cause might be the too small tile size used for training. Indeed, the DeepLabv2 architecture is built to learn large receptive fields thanks to its dilated convolutions. Using 636$\times{}$636 tiles may hinder the learning process by forcing the model to learn smaller receptive fields than it was designed for in order not to suffer from border effects.

\subsection{Transferability to other domains}
\label{sec:generalization_of_trained_models}
Pre-training not only helps improving performances on the source domain but also brings generalizability to the trained models. Table \ref{tab:100-200ep} provides the obtained mIoU on the T2 domain when applying the best source domain trained models after 100 and 200 epochs of fine-tuning. The 200 epoch results were already contained in Table \ref{tab:target2}. Generally, pre-trained models perform better, especially with NanoSEM. Interestingly, when pre-training with ImageNet, models’ generalization power decrease between 100 and 200 epochs. It can be assumed that, during prolonged training, weights are tweaked such that very dataset-specific features are progressively replacing general ones. While after 100 epochs, using NanoSEM pre-training achieves the best results, the difference between ImageNet and the two-stage NanoSEM after 200 epochs is marginal. However, quantifying and visualizing this phenomenon of increasing model specialization through prolonged training remains an open question.

Contrary to T1, the basic augmentation pipeline consistently outperforms the extended one for T2. This poor domain generalizability (Exp.\# T2.3, T2.6, T2.9) suggests that models were rendered invariant to some task-relevant features of T2 when trained with the extended pipeline. It should be emphasized again that this pipeline was optimized for the source domain, which exhibits a substantially wider domain gap with T2 compared to T1. Extended data augmentation causing a drop of generalizability has previously been observed in\cite{Thomas2020}.

\begin{table}[ht]
\caption{Source domain-trained models' performance on the target 2 domain (cf. Table \ref{tab:datasets}) when fine-tuning for 100 or 200 epochs. Indices refer to training epochs. Provided mIoUs are averaged over the five folds used for cross-validation. NanoSEM refers to the two-step pre-training introduced in Section \ref{subsec:PT}. Augmentation pipelines are detailed in the Supplemental.}

	\centering
		\begin{tabularx}{\linewidth}{
	p{\dimexpr.15\linewidth-2\tabcolsep-1.3333\arrayrulewidth}
	p{\dimexpr.14\linewidth-2\tabcolsep-1.3333\arrayrulewidth}
    p{\dimexpr.11\linewidth-2\tabcolsep-1.3333\arrayrulewidth}
    p{\dimexpr.07\linewidth-2\tabcolsep-1.3333\arrayrulewidth}
    p{\dimexpr.14\linewidth-2\tabcolsep-1.3333\arrayrulewidth}
    p{\dimexpr.13\linewidth-2\tabcolsep-1.3333\arrayrulewidth}
    p{\dimexpr.14\linewidth-2\tabcolsep-1.3333\arrayrulewidth}
    p{\dimexpr.13\linewidth-2\tabcolsep-1.3333\arrayrulewidth}
	}
	 \textbf{Model} & \textbf{Pre-training} & \textbf{Augment.} & \textbf{Exp.\#} &
	 \textbf{mIoU$_{100}$ [\%]} & \textbf{RDT$_{100}$ [\%]}
	 &
	 \textbf{mIoU$_{200}$ [\%]} & \textbf{RDT$_{200}$ [\%]}\\
	\thickhline
    \multirow{9}{*}{VGG16 U-Net} & \multirow{3}{*}{Random init.} & --- & T2.1 & 50.1 ± 3.6 & -33.7 & 51.4 ± 4.9 & -32.1\\
    \cline{3-8} & & basic & T2.2 & 56.4 ± 4.0 & -27.0 & 57.6 ± 4.9 & -26.0\\
    \cline{3-8} & & extended & T2.3 & 49.1 ± 8.5 & -34.2 & 52.5 ± 8.5 & -31.5\\       
    \cline{2-8} & \multirow{3}{*}{ImageNet} & --- & T2.4 & 57.3 ± 1.6 & -24.2 & 58.1 ± 2.4 & -23.7\\
    \cline{3-8} & & basic & T2.5 & 63.4 ± 1.3 & -19.5 & 63.8 ± 1.1 & -19.2\\
    \cline{3-8} & & extended & T2.6 & 59.9 ± 5.2 & -24.1 & 52.8 ± 11.8 & -33.2\\
    \cline{2-8} & \multirow{3}{*}{NanoSEM} & --- & T2.7 & \textbf{60.0 ± 2.6} & \textbf{-21.3} & 59.7 ± 3.0 & -22.3\\
    \cline{3-8} & & basic & T2.8 & \textbf{63.9 ± 2.4} & \textbf{-19.1} & 61.2 ± 3.7 & -22.7\\
    \cline{3-8} & & extended & T2.9 & \textbf{63.2 ± 1.9} & \textbf{-20.5} & 60.3 ± 3.3 & -24.8\\   
	\thickhline
\end{tabularx}
\label{tab:100-200ep}
\end{table}

Concerning the UDA framework, the obtained results are very encouraging. For T1, it appeared that due to the minimal domain shift with respect to the source, transferring source-trained models was already performing satisfactorily (cf. experiments T1.6, T1.9). Hence, this problem posed to UDA is not overly challenging. The DG models' achieved mIoUs on this target domain even exceed the source mIoUs. Presumably, this can be attributed to the additional parallel features introduced by the subgrain boundaries (see Figure \ref{fig:datasets}), rendering the prediction easier. This was verified using the GradCAM network visualization technique, which computes the network gradients at a specific layer with respect to a target class and thereby estimates pixel-wise activation. For more information, we refer to \cite{GradCAM}.

\begin{figure}[htbp]
	\centering
	\footnotesize
	\includegraphics[width=\linewidth]{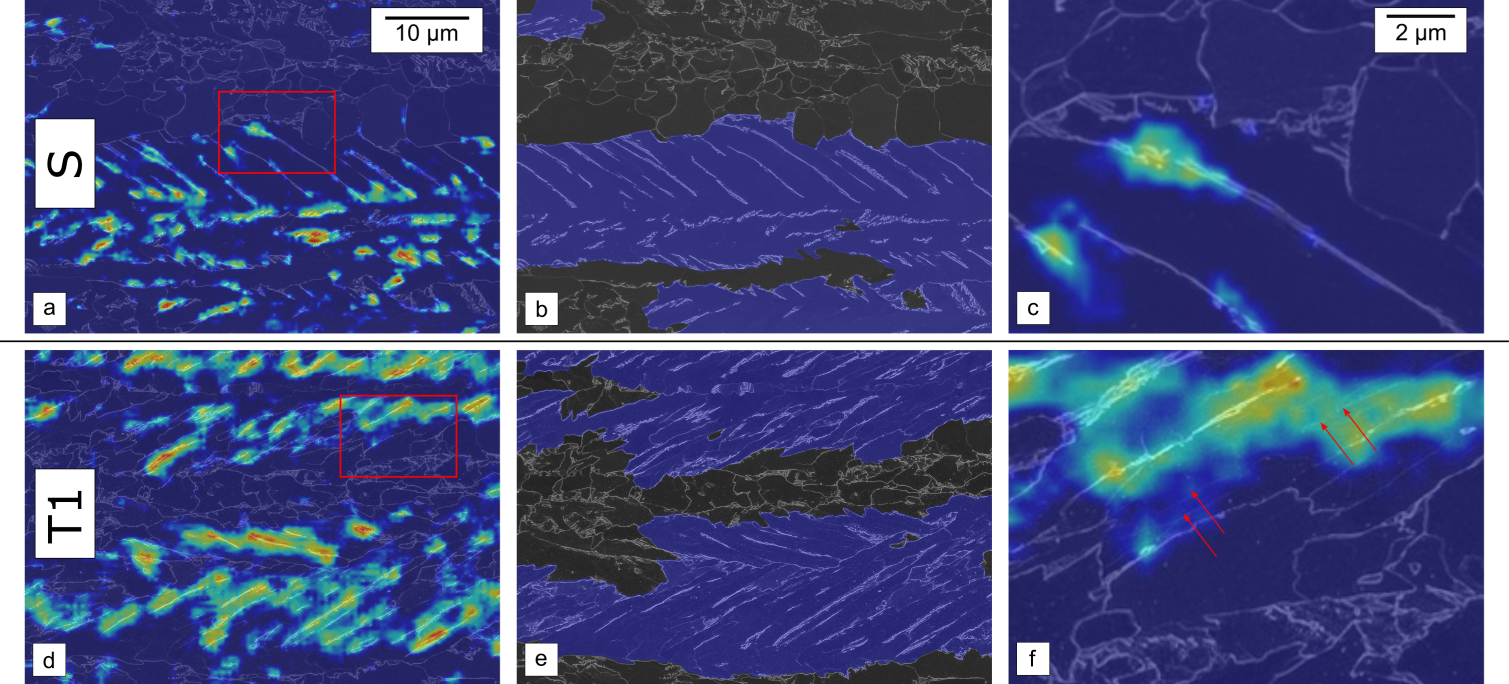}
	\caption{Network visualizations of the ResNet-101 DeepLabv2 model (supervised) on the source (a--c) and target 1 (d--f) domain with lath-shaped bainite labels (b) and (e) given as reference. The heat maps indicate regions that were taken into consideration at layer 3\_16 of the ResNet-101\cite{he2015deep} to classify the lath-shaped bainite regions. In the detail views (c), (f), which are the same as in Figure \ref{fig:datasets}, heat maps in the target 1 domain are more extensive and additionally incorporate subgrain boundaries (red arrow annotations).}
	\label{fig:GradCAM}
\end{figure}

In Figure \ref{fig:GradCAM} this technique has been applied to the source and T1 domain to determine regions that were deemed important by the (same) ResNet-101 model (trained on source). It is clear that the activation is more extensive in the T1 domain and additionally involves the subgrain boundaries inside the laths (red arrow annotations). This supports the theory that these additional features induced by Nital etching are beneficial for the model. In our prior study, we discovered that image downscaling for the source domain culminates in a performance increase since the pixel gap between carbides at lath boundaries is reduced, and information loss is minimal \cite{Durmaz2021}. Therefore, the parallelism of these features can be assessed at earlier network layers. The GradCAM results on layer 3\_16 indicate that the hierarchical microstructure and internal subgrain boundary features revealed in T1 can help to bridge the otherwise feature-sparse bainitic ferrite regions in the source domain to improve learning. Note that the activation in Figure \ref{fig:GradCAM}c is high where parallel carbide films are in close vicinity. These parallel carbide films being decisive features indicates that these trained models' performance could be compromised when evaluated on cross-sections in rolling or normal direction due to their distinct microstructural patterns \cite{thornton2008three}. Moreover, considering the small test sets, it can not be excluded that the T1 test images potentially being easier to predict on average contributes to the better T1 performance. 
Overall the UDA framework gave a 2.5\% mIoU boost on this target domain (T1.11 compared to T1.10). Furthermore, training the UDA framework with T1 as the target domain gave models that perform better on the \textit{source domain} with 79.7\% mIoU, granting a 0.3\% boost compared to experiment S.10. Similarly, such small domain gaps led to the same observation in the context of urban images \cite{bolte2019unsupervised}. 

On the other hand, T2 and T3 have broader gaps with respect to the source domain due to stronger etching and different imaging modalities, respectively. Despite the large domain gap of the T3 dataset with the source one, UDA performed substantially better on this dataset than on T2, culminating in a 23.6\% mIoU improvement over the DG experiments, which is mirrored in Figure \ref{fig:before-after_Target3}. As a reference, fully supervised training on T3 presented in our last paper \cite{Durmaz2021} achieved 79\% mIoU, showing that employing UDA results in a 6\% drop only, despite not relying on target labels. This result with respect to bridging modalities is promising and in line with literature where domain adaptation in the medical field was successfully applied to transition between computer tomography and magnetic resonance imaging \cite{Dou2018}. Note that UDA models trained with T2 and T3 datasets scored 78,8\% and 75,8\% mIoU on the \textit{source domain}, respectively, falling short compared to the fully-supervised training (Exp.\# S.10: mIoU = 79.4\%). Indeed, UDA reaches a compromise between source and target, which is detrimental to the source domain when large domain gaps with the target are involved. This observation confirms that T1, T2, and T3 are gradually increasing the domain shift with respect to the source domain.

The difference in UDA performance on the target domain between T2 and T3 could be attributed to the 5$\times{}$ larger data amount available for the latter set (see Table \ref{tab:datasets}), where 48 unannotated training tiles for T2 might be insufficient. Another reason could be the phase fraction of the T2 dataset, which is substantially lower than the source dataset. This assumption will be discussed in the following Section \ref{sec:limits_UDA}. Lastly, we consider it unlikely that the SEMs different distortion and noise level characteristics (see Section \ref{sec:methods}) are causing this difference since the UDA framework can cope with different modalities and corresponding data augmentations were applied.

Additionally, to improve over the individually trained UDA models, we also implemented a bagging strategy. This consists in averaging the predictions of multiple models to give an improved segmentation. In our case, we use the models trained with the five different folds and achieve 85.9\%, 70,2\% and 75.0\% mIoU, which results in 1.2\%, 2.9\% and 1.7\% increase over the best results presented in the T1, T2, and T3 result tables, respectively. The larger improvement for T2 can be attributed to its comparatively weak individual classifiers, making the bagging paradigm relatively more profitable. 

Lastly, as opposed to DG models, which are frequently biased towards a class (cf. Figures \ref{fig:before-after_Target1}, \ref{fig:before-after_Target2}), UDA leads to more balanced models. Consequently, this characteristic improves the estimation of phase fractions or other metrics, which do not require full details of the segmentation mask. As an example, predicted phase fractions on the different target datasets with DG and UDA are reported in Table \ref{tab:PhaseFractions}. It is evident that UDA improves phase fraction estimation in all cases and additionally reduces the scatter substantially (e.g., by a factor of 10 for T3). The large observed scatter of DG models for T2 and T3 predictions can be explained by the different training folds, each leading to skewed predictions in favor of either background or foreground class. Using UDA systematically reduced the skewed behavior and gave models that are only slightly biased towards the background class. This suggests that UDA-based models, when applied to the target domains, misinterpret some foreground class features. One potential cause could be incompletely bridged gaps with the source domain. Another possibility for this consistent lath-bainite underestimation could be the labeling process. Unlike the target domains, the source was labeled with supporting EBSD images, which could conceivably result in different annotation patterns. Lastly, one should recall that the fairly small test set size renders the results very sensitive concerning any labeling inconsistencies. For these reasons, the impact of intra-rater reliability is presumably elevated.

\begin{table}[ht]
\caption{Estimated phase fractions on the test sets of the different target domains with DG and UDA models. Values are averaged over the five models trained on the different folds and standard deviations are provided.}

	\centering
		\begin{tabularx}{\linewidth}{
	p{\dimexpr.28\linewidth-2\tabcolsep-1.3333\arrayrulewidth}
	p{\dimexpr.28\linewidth-2\tabcolsep-1.3333\arrayrulewidth}
    p{\dimexpr.28\linewidth-2\tabcolsep-1.3333\arrayrulewidth}
    p{\dimexpr.15\linewidth-2\tabcolsep-1.3333\arrayrulewidth}
	}
	 \textbf{Target domain} & \textbf{$\phi_{DG}$ [\%]}  & \textbf{$\phi_{UDA}$ [\%]} & \textbf{$\phi_{label}$ [\%]} \\
	\thickhline
	Target 1 (T1) & 39.8 ± 1.0 & 41.3 ± 0.9 & 43.6 \\
	\hline
	Target 2 (T2) & 48.4 ± 10.7 & 44.6 ± 4.9 & 46.7 \\
	\hline
	Target 3 (T3) & 51.3 ± 21.8 & 42.8 ± 2.6 & 44.6 \\
	\thickhline
\end{tabularx}
\label{tab:PhaseFractions}
\end{table}

\subsection{Limitations and potentials of the implemented unsupervised domain adaptation framework}
\label{sec:limits_UDA}

Firstly, it has to be emphasized that adversarial-based frameworks such as AdaptSegNet suffer from training instability, rendering them rather laborious to tune and affecting training repeatability. Facing these training pitfalls while working with low quantity data hampers the deduction of relationships from ablation studies and hyperparameter studies. Furthermore, given the implementations and hardware at hand, typical UDA training runs for six hours, whereas a DeepLabv2 supervised fine-tuning takes only 45 minutes.

Another constraint was witnessed when trying to bridge the gap between SEM and LOM data (T3). Indeed, the learning process failed when feeding the model with bright-field LOM tiles, motivating us to invert their pixel values. As the segmentation architecture shares weights between source and target data, it might be difficult to learn filters, which perform well on both modalities while keeping internal features independent of the originating domain.

Moreover, this AdaptSegNet \cite{tsai2020learning} framework is built on the strong prior assumption that the source and target datasets are sharing the same label space distribution. This poses a boundary condition for the segmentation model to give good predictions and fool the discriminator simultaneously.
In case of pronounced label space deviations between the source and target datasets, the discriminator should hypothetically quickly learn how to differentiate the segmentation masks, hampering the transfer learning process. Such a label distribution shift could be due to different phase morphology or phase fractions. For this purpose, the phase fractions are provided in Table \ref{tab:datasets}. Aside from generally low data quantity in the T2 domain, the lath bainite phase was not oversampled during image acquisition as opposed to the other domains. Therefore, selecting a suitable training tile subset to match the source domains' 53\% mean lath-bainite phase fraction was unfeasible. Taking the phase fraction histograms of the training tiles (based on expert-reviewed pseudo-labels) into consideration, the discriminator seemingly should learn the tendency that the images from the T2 domain show generally a smaller lath-bainite content. Initially, we considered this to be the primary reason for UDA being more beneficial for T3 compared to T2. However, an additional experiment where we varied the lath-bainite phase fraction of the LOM training dataset provided to the UDA framework invalidated this hypothesis. Specifically, we sampled another target 3$^v$ train set with a lower mean phase fraction ($\phi_{train,3^v} = 0.28$ similar to T2) and trained a model with it in the UDA framework. A common test set exempt from both LOM training sets was created for testing purposes consisting of 12 test images with a mean phase fraction of 40\%. This LOM test set phase fraction was chosen to be the average between the training phase fractions such that the influence of target train-test shifts could be excluded and domain phase fraction shifts during training could be investigated. The results showed that models trained with T3 reached 78.0 ± 1.0 mIoU, whereas those trained with T3$^v$ scored 77.4 ± 1.8. Note that these results deviate from the results provided in Table \ref{tab:target3} due to the distinct test set. This suggests that distinct phase fraction training data of the different domains does not hamper the UDA training. One thing to underline is that the discriminator receives predictions rather than actual annotations, which complicates the distinction based on phase fraction histogram separability. Nevertheless, the robustness of this adversarial process concerning space label non-conformity is auspicious for materials science tasks. For instance, this appears promising for generalizing to different alloys or processing routes, as the phase topology and morphology then can be altered significantly between the source and target sets. Alternatively, if problems due to too different phase distributions were to arise, these could be overcome by feeding the discriminator with tile sub-patches selected based on pseudo-labels to balance both sets' apparent tile phase fractions artificially.

We consider the employed UDA model a good trade-off between complexity and reached performance, and therefore an excellent introductory framework for the material science community. However, in view of the fast-paced ML research, it has been outperformed on the GTA5-to-cityscape task. Several improvements have been published over the past three years, most of the time using the work of Tsai et al.\cite{tsai2020learning} as a reference and starting point \cite{pan2020unsupervised, tranheden2020dacs, vu2018advent, yang2020fda, Yu_Zhang_Dong_Hu_Dong_Zhang_2021}. All these studies rely on the GTA5-to-Cityscapes reference task, and some of them exploit specific characteristics of these datasets \cite{tranheden2020dacs}. Therefore, the approaches in these works are not directly applicable for our binary segmentation task but potentially relevant for other material science tasks. Nevertheless, some models could potentially improve over AdaptSegNet in our setting. One example is the ADVENT model of Vu et al. \cite{vu2018advent}, which makes use of entropy maps instead of segmentation maps as input for the discriminator. It encourages entropy minimization in the target domain by matching the source and target entropy distributions. This entropy minimization paradigm is borrowed from semi-supervised learning. Also inspired by semi-supervised techniques, Pan et al. \cite{pan2020unsupervised} implement the AdaptSegNet framework with an extra pseudo-labeling step. The easiest-to-predict half of the target data is pseudo-labeled in a first training iteration and then utilized as "source" domain data for a second training, using the rest of the target data as the target domain. This is motivated by the intra-domain variance in the target domain. In our case, this approach would probably not be overly beneficial as our intra-domain variance has been reduced to the minimum by repeatable data acquisition. Potentially, such an approach can take the minor emerging intra-domain variance from grain morphology differences due to imaging at different locations on the rolled sheet cross-sections into account. Recently, Yu et al. published an improvement of AdaptSegNet\cite{tsai2020learning} including an attention mechanism in order to focus domain adaptation on the parts of the images that are the most difficult to transfer from the source to the target.
While the work at hand has focused on adversarial UDA techniques only, promising style transfer GAN approaches are also good candidates for UDA methods and are currently an active research topic.

Whether to employ UDA for training a model depends on three criteria. First, the effort associated with labeling source and target domain data needs to be considered since UDA avoids this cost for the target. For instance, UDA is especially favorable when synthetic source data (e.g., simulation data) with inherent labels and expensive target annotations are concerned. Second, the features contained in the source and target input data determine the attainable annotation accuracy. A setting where significantly more precise labels can be obtained in a source domain compared to the target proves beneficial to UDA. Thus, assuming comparatively poor-quality target labels, the performance gap between a UDA training and the direct supervised training on the target domain diminishes. The transfer from SEM to LOM provides a good example as SEM image features not only render the phase annotation process easier but also the bainite sub-class differentiation possible in the first place. Therefore, considering that a SEM acquires data substantially slower and is not affordable for every research laboratory, training a UDA model with external annotated SEM data to transfer to LOM can increase accessibility to high-quality models and potentially even enable specific tasks. Lastly, the domain gap to bridge has to remain maintainable. Source and target domains need to share enough common image features for the model to learn a domain-independent representation of the data. While this work gives first insights into the UDA scope of application, its precise limits still need to be explored.

UDA could alleviate the demand for expensive annotations, considering the variety of characterization methodologies and materials utilized in materials science. For instance, can surface-sensitive SE2 SEM images be used in a UDA setting to transfer to topography mapping techniques such as atomic force microscopy? Or accordingly, can an alloy quenched with cooling rate A be used to transfer to a cooling rate B? These are essential tasks to experimentally confirm computationally optimized microstructures, given the materials design acceleration we currently undergo.
\section{Conclusion and Outlook}
\label{sec:conclusion_and_outlook}

Amidst others \cite{gonthier2020analysis, Romero2020}, this study validates the benefit of pre-training when a low-data scenario is encountered. In this context, we show that a two-stage pre-training using ImageNet and an in-field SEM dataset also improves the generalizability of the trained models across domains. From the results, it is evident that models learn and forget relations, and generalizability is sacrificed for specificity upon prolonged training. The success of pre-training and fine-tuning motivates the demand for publicly available datasets in the material science field.

Concerning model transferability, this works' significant contribution lies in the successful application of Tsai et al.'s UDA approach \cite{tsai2020learning} with small datasets, making this technique promising for many applications. Substantial improvements in performance on the target datasets were observed despite only providing few tens of unlabeled micrographs. Even modality transfers from SEM to LOM could be facilitated successfully with such data. The UDA frameworks' insensitivity with respect to different phase fractions in source and target domains yields hope to enable generalization across different alloys and heat treatments. Considering the increasing image acquisition rates and automation, the discrepancy between labeled and unlabeled data quantities is very likely to dramatically increase in the future, raising interest for unsupervised-learning methods (e.g. UDA). Indeed, the availability and expense of annotation processes in material science are impediments, that UDA can evidently help to overcome. This applies especially when source domain data is substantially cheaper to annotate, shows relevant features for labeling precisely, and when domain gaps to bridge are narrow enough (comparable features between the domains).

Lastly, efforts have been made over the past years to infuse image classifiers with semantics and structured knowledge \cite{marino2016more, wang2018zero}, to alleviate the restrictions with respect to data efficiency and generalizability across datasets. These models utilize knowledge modeling in the form of ontologies and derived knowledge graphs as well as graph neural networks to introduce reasoning capability to models. As materials science currently undergoes the digital transformation in numerous large scale projects \cite{MGI, o2019materials, MD}, resulting knowledge graphs will not only increase the findability of potential source training data but can, along with the techniques presented here, potentially shape the future of generalizable learning to account for materials diversity.

\section*{Data availability}
The datasets generated during and/or analysed during the current study are not publicly available because they are part of an ongoing study and subject to third party (AG der Dillinger Hüttenwerke) restrictions.

\bibliography{bib_UDA.bib}



\section*{Acknowledgements}
We express our appreciation towards project DEAL for provided Open Access funding.

\section*{Author contributions statement}
Conceptualization: A.D.; Data curation: A.D., A.G, M.M; Formal Analysis: A.G. ; Investigation: A.D., A.G.  Methodology: A.D., A.G., A.T.; Project administration: A.D., P.K.; Resources: A.D., C.E. D.B; Software: A.G., A.T.; Supervision: A.D., C.E., P.K.; Visualization: A.D., A.G.; Writing – original draft: A.D., A.G., C.E., M.M., P.K.; Writing – review \& editing: all authors;

\section*{Additional information}

\textbf{Competing interests} The authors declare no competing interests.

\end{document}